\documentclass[epjST]{svjour}
\usepackage{dcolumn}
\usepackage{graphicx,color}
\usepackage{amsmath,amssymb,bbm,bm}
\usepackage{csquotes,mathrsfs}

\newcommand{\ket}[1]{|{#1}\rangle  }
\newcommand{\braket}[2]{\langle{#1}| {#2}\rangle}
\newcommand{\ketbra}[2]{\vert {#1} \rangle \langle{#2}\vert}

\begin{document}
\title{Probing Ultrastrong Light-Matter Coupling in Open Quantum Systems}
\author{A. Ridolfo\inst{1,3}\and J. Rajendran\inst{1,3} \and L. Giannelli\inst{1,3} \and E. Paladino\inst{1,2,3} \and G. Falci\inst{1,2,3}\fnmsep\thanks{\email{giuseppe.falci@unict.it}} }
\institute{Dipartimento di Fisica e Astronomia "Ettore Majorana", Universit\'a di Catania, Via S. Sofia 64, 95123, Catania, Italy \and CNR-IMM, UoS Universit\`a, 95123, Catania, Italy \and INFN, Sez. Catania, 95123, Catania, Italy}
\abstract{Dynamically probing systems of ultrastrongly coupled light and matter by advanced coherent control has been recently proposed as a unique tool for detecting peculiar quantum features of this regime. Coherence allows in principle on-demand conversion of virtual photons dressing the entangled eigenstates of the system to real ones, with unitary efficiency and remarkable robustness. Here we study this effect in the presence of decoherence, showing that also in far from ideal regimes is it possible to probe such peculiar features.} 
\maketitle

\section{Introduction}
Light-matter interaction is a fundamental building block of Nature leading to countless applications, whose scope broadened in recent years 
with the advent of quantum technologies~\cite{kb:06-harocheraimond}. Circuit-QED solid-state systems  are one of the forefront platforms for quantum hardware~\cite{kr:08-schoelkopf-nature-wiring} where besides applications, fundamental physics from measurement theory~\cite{ka:04-wallraff-superqubit} to quantum thermodynamics~\cite{ka:18-distefanopaternostro-prb-measurethermo} and quantum communication~\cite{ka:09-benentifalci-prl-memorych} can be studied. The basic physics is described by the Rabi model~\cite{ka:137-rabi-pr-rabimodel}, where a two-level atom interacts with a single quantized oscillating mode. In the strong coupling regime (SC) the system exhibits exquisitely quantum dynamics, conserving the number of excitations to an enormous degree of accuracy, this approximate symmetry breaks down in the ultrastrong coupling (USC) regime~\cite{ka:05-ciuti-prb-intersubbandpolariton}.  Systems in this regime have been demonstrated in the last few years in circuit-QED architectures of solid-state artificial atoms (AA), THz metamaterials, intersubband polaritons and other physical systems~\cite{kr:19-kokchumnori-natrevphys-usc,kr:19-forndiazsolano-rmp-usc}. USC systems have been subject of extensive theoretical investigation predicting novel and outstanding non-perturbative effects~\cite{kr:19-kokchumnori-natrevphys-usc,kr:19-forndiazsolano-rmp-usc,ka:12-ridolfo-prl-photonblock,ka:13-ridolfo-prl-nonclassicalradiation}. Amongst them, it has been pointed out that non-conservation of the excitation number implies that the ground state $\ket{\Phi_0}$ of the Rabi model in the USC regime contains virtual photons. Their detectability has been the subject of intense theoretical investigation~\cite{ka:12-carusotto-prl-3levusc,ka:13-stassisavasta-prl-USCSEP,ka:14-huanglaw-pra-uscraman,ka:17-falci-fortphys-fqmt,ka:17-distefanosavastanori-njp-stimemission,ka:19-falci-scirep-usc,19-deliberato-pra-quantumvacuum} but experimental demonstrations are still lacking. In order to understand and fill this gap the very recent work~\cite{ka:19-falci-scirep-usc} addressed key issues on hardware design and control, showing how to demonstrate efficient and unambiguous photon pair conversion, amplified using an advanced coherent control protocol analogous to stimulated Raman adiabatic passage (STIRAP)~\cite{kr:17-vitanovbergmann-rmp}, which has been proposed and demonstrated in 
AAs~\cite{ka:09-siebrafalci-prb,ka:15-distefano-prb-cstirap,ka:16-kumarparaoanu-natcomm-stirap,ka:16-xuhanzhao-natcomm-ladderstirap,ka:16-vepsalainen-photonics-squtrit}. In this work we focus on the evaluating the impact of decoherence processes in AAs~\cite{kr:14-paladino-rmp} on the dynamics of the ideally quantum-controlled systems.

\section{Coherent amplification of photon pair production}
The simple version of the coherent amplification protocol technique was proposed in Ref.~\cite{ka:17-falci-fortphys-fqmt}.
A three-level atom (basis $\{\ket{u},\ket{g},\ket{e}\}$) is ultrastrongly coupled to a quantized mode with oscillation frequency $\omega_c$, resonant with the $e\!.g\!$ transition energy $\varepsilon$. The  Hamiltonian is ($\hbar =1$)
\begin{equation}
H = \varepsilon \, | e \rangle \langle e | - \varepsilon^\prime \, | u \rangle \langle u | + \omega_c \,a^\dagger a + \lambda (a+a^\dagger)(| e \rangle \langle g | + | g \rangle \langle e |) 
\label{eq:H-rabi-model}
\end{equation}
where $\varepsilon^\prime$ is the atomic $u\!-\!g$ splitting, $a$ and $a^{\dagger}$ are the photon annihilation and creation operators, the eigenstates of $a^\dagger a$ forming the basis of the Fock states $\{\ket{n}\}$. Here $\lambda$ is the coupling between the mode and the $e\!-\!g$ transition, whereas $\ket{u}$ is uncoupled (as for $\epsilon^\prime-\omega_c \gg \lambda$). The so called diamagnetic term~\cite{kr:19-forndiazsolano-rmp-usc,kr:19-kokchumnori-natrevphys-usc} is implicitly accounted for by renormalized parameters, but it plays a negligible role in the regime of couplings considered in this work.  
Eigenstates of $H$ are partitioned in two sets: (1) normalized eigenstates of the two-level Rabi model, denoted by $\ket{\Phi_j}=\sum_n c_{jn} \ket{ng} + d_{jn} \ket{ne}$, with eigenvalue $E_j$ and amplitudes $\{c_{jn},d_{jn}\}$; in the USC regime they are strongly entangled atom-mode states, dressed by virtual photon pairs; (2) factorized states $\ket{n u}$, with energy $\varepsilon^{\prime} + n \, \omega _c$, 
which will be used as ancillas. Extra terms may trigger transitions between these subspaces. In particular atomic $u-g$ couplings may determine 
the conversion of virtual photons of the Rabi subspace in real photons of the ancillary subspace, where they can be detected. At resonance, $\omega_c = \epsilon$ virtual photons in the (false) Rabi vacuum $\ket{\Phi_0}$ are directly converted to real photon pairs, the main channel involving the target state $\{\ket{u\, 2}\}$, with corresponding amplitude $c_{0\,2}(\lambda/\omega_c)$. Since this latter is large enough only in the USC regime, the detection of photon pairs, while the AA is found in $\ket{u}$, provides a unique dynamical signature of USC.

\begin{figure}[t!]\centering\resizebox{0.95\columnwidth}{!}{\includegraphics{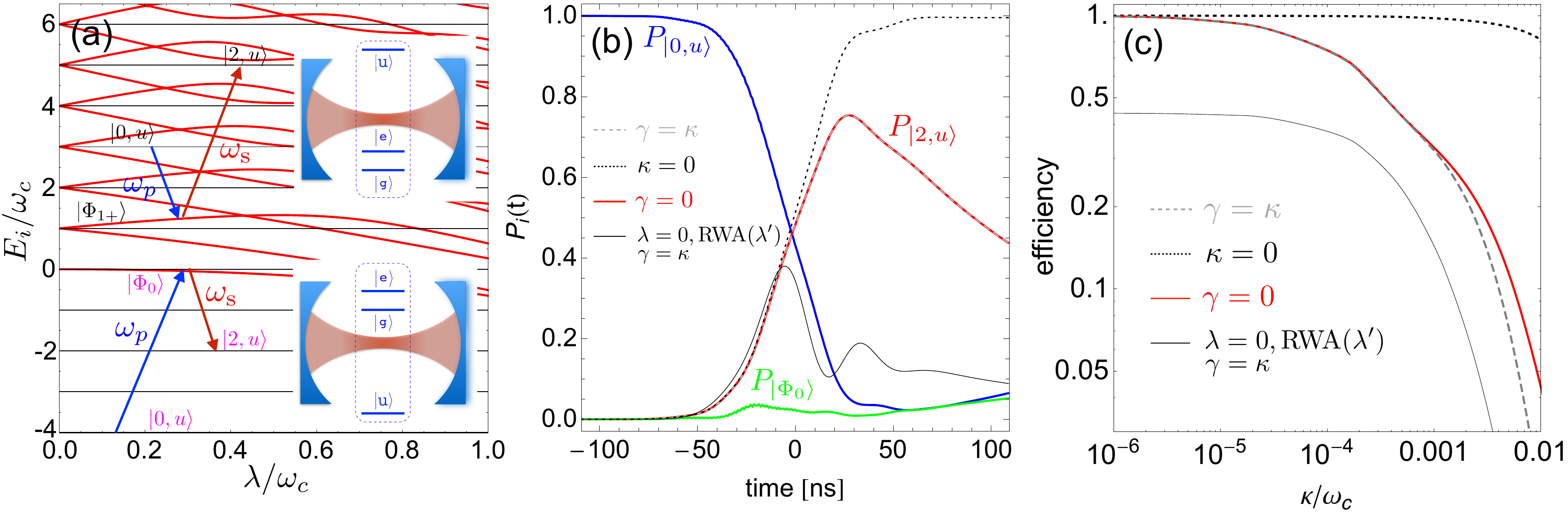}}
\caption{(a): level scheme of $H$ as function of the coupling strength $\lambda$ for $\epsilon^{\prime} = 4\epsilon$ and $\epsilon=\omega_{c}$. States $|0,u\rangle$, $|2,u\rangle$ and $|\Phi_{0}\rangle$ are the minimal set for $\Lambda$ STIRAP and the blue and red arrows represent the resonant pump and Stokes fields respectively. (b): population histories of $\Lambda$ STIRAP manifold calculated for $\lambda/\omega_{c} = 0.5$, $\lambda^{\prime} = 0$ (colored lines), ${\cal W}_{s}/\omega_{c} = 0.1$, ${\cal W}_{p}/{\cal W}_{s} = 0.0972$, $\kappa/\omega_{c} = 10^{-4}$, $\gamma=0$, $\epsilon^{\prime} = 4\epsilon$, $\epsilon = \omega_{c} = 6 \, \text{GHz}$, $T = 54.6 \, \text{ns}$ and $\tau = 0.6 T$. Moreover, $P_{\ket{2,u}}(t)$ is calculated for $\kappa = 0$ and $\gamma/\omega_{c} = 10^{-4}$ (black dotted), $\gamma = \kappa =  10^{-4} \omega_{c}$ (gray-dashed). Population transfer due to a corotating stray coupling  $\lambda^{\prime}/\omega_{c} = 0.5$ and $\lambda = 0$ for $\gamma=\kappa=10^{-4}\omega_{c}$ is also shown (black solid thin). (c): efficiency of the protocol as function of $\kappa$ with the same parameters of (b).
\label{fig:0}}    
\end{figure}

The above photon pair production channel could be amplified, and thus made detectable, by using coherent control. The atom is driven by two-tone control field in $\Lambda$ configuration described by the Hamiltonian (see Fig.\ref{fig:0}a)
\begin{equation}
H_{\rm C}^{\Lambda}(t) = W(t) (|u\rangle\langle g| + |g\rangle\langle u|)
\quad ; \quad 
W(t) = \sum_{k=p,s} \mathscr{W}_k(t) \, \cos(\omega_k t) 
\end{equation}
We take $\omega_p \approx E_0 - \varepsilon_u$, $\omega_s \approx \omega_p -2 \omega_c$, Gaussian envelopes ${\mathscr W}_{s/p}(t) = {\mathscr W}_{s,p} \exp[-(t \pm \tau)^2/T^2]$ with width $T$, delay $\tau>0$  and Stokes/pump peak field amplitude ${\mathscr W}_{s/p}$~\cite{ka:17-falci-fortphys-fqmt,kr:17-vitanovbergmann-rmp}.  

This protocol yields non-vanishing coherent population transfer $\ket{0u} \to \ket{2 u}$ only if $\braket{\Phi_0}{H_c|nu} \neq 0$, i.e. when 
$\ket{\Phi_0}$ contains pairs of virtual photons. In the  regime $0.2 \lesssim \lambda/\omega_c \lesssim 0.5$, which is very interesting for experiments the dynamics can be conveniently visualized as a $\Lambda$-STIRAP population transfer in the \textquote{STIRAP subspace} $\{\ket{u\,0},\ket{u\,2}, \ket{\Phi_0}\}$ and control may be tailored to yield $\sim 100\%$ probability to find eventually two photons~\cite{ka:19-ridolfofalci-epj-usc}, the AA always remaining in the state $\ket{u}$.    

We remark that the implementation of dynamical detection schemes in real systems poses severe constraints and challenges. This has been exhaustively discussed in the works~\cite{ka:19-falci-scirep-usc,ka:19-ridolfofalci-epj-usc}, where detection of virtual photon pairs was studied in superconducting systems using the flux-qubit design and advanced control~\cite{ka:16-distefano-pra-twoplusone}. Here we focus on the single issue of evaluating the impact of decoherence on the dynamics of the ideally quantum-controlled systems.

\section{Partially coherent photon pair conversion}

We stress that for the detection of USC it would be sufficient to monitor an increase of the population of the Fock states $\ket{n\ge 2}$ during part of the protocol. Some transient population of the intermediate state is also tolerable, softening the adiabaticity requirement. Decoherence times $T_\phi\sim T$ can also be tolerated~\cite{ka:13-falci-prb-stirapcpb} since at worst efficiency of the USC-selective channel would be larger than $30\%$. This opens perspectives also for semiconducting structures, where USC-selective $\Lambda$-STIRAP could be observed with some progress in techniques for detecting excess THz photons. Therefore we address the problem of observability of photon pair conversion when coherence is partially lost. 

\subsection{Modeling decoherence}
The open quantum system dynamics for $\Lambda$ STIRAP, is studied considering Markovian decay processes of both the cavity and the AA, accounted for by the following Lindblad equation 
\begin{equation}
\label{eq:lindblad-lambda}
\dot{\rho}(t) = {\rm i}[\rho, H + H_{\rm C}^{\Lambda}(t)] + {\cal L}_{\rm cav}[\rho] + {\cal L}_{\rm atom}[\rho]
\end{equation}
The dissipator ${\cal L}_{\rm cav}[\rho] = \frac{\kappa}{2} (2 a \rho a^{\dagger} - a^{\dagger} a \rho - \rho a^{\dagger} a)$ describes losses of the cavity mode with rate $\kappa$. Spontaneous decay of the three-level atom is modeled by the dissipator  
${\cal L}_{\rm atom}[\rho] = \frac{\gamma}{2} (2 \sigma_{\rm ug} \rho \sigma_{\rm ug}^\dagger - \sigma_{\rm gg} \rho - \rho \sigma_{\rm gg})
+ \frac{\gamma}{2} (2 \sigma_{\rm ge} \rho \sigma_{\rm ge}^\dagger - \sigma_{\rm ee} \rho - \rho \sigma_{\rm ee})$ 
where $\sigma_{\rm ug}=| u \rangle \langle g| $, $\sigma_{\rm gg}=| g \rangle \langle g|$,
$\sigma_{\rm ge}=| g \rangle \langle e| $ and $\sigma_{\rm ee}=| e \rangle \langle e|$. 
This is a minimal model aiming to understand the impact of leading processes, assuming suitable external conditions.
Temperature much smaller than $\omega_c$  is assumed as well as and not so strong driving, in order to neglect atomic absorption and stimulated emission.  We discard $e-u$ transitions, having in mind a three-level AA implemented by a devices biased at a symmetry point, which enforces parity selection rules. Symmetries also cancel at lowest order pure dephasing due to low-frequency noise both of the AA~\cite{ka:03-paladino-advssp-decoherence,ka:08-paladino-prb-coherimp,ka:10-paladino-prb-opt2qubit,kr:14-paladino-rmp} and the cavity.
We moreover assume that the environments has power spectrum slowly varying in a frequency interval $\delta \omega \sim \lambda$ around 
$\omega = \omega_c$, so that rates are weakly dependent on $\lambda$.


\subsection{Results for the $\Lambda$ configuration}
The dynamics for $\Lambda$ scheme is carried out by numerically solving Eq.(\ref{eq:lindblad-lambda}). 
An example of population histories for $\kappa\neq0$ and $\gamma=0$ is shown in Fig.\ref{fig:0}b, colored lines, which exhibit the increase of $P_{\ket{2u}}$ due to USC while $P_{\ket{0u}}$ and $P_{\ket{\Phi_0}}$ remain small. Notice that population eventually leaks outside the \textquote{STIRAP subspace} due to photon decay $\ket{2 u} \to \ket{1 u}$ in the cavity whose effect is studied in Fig.Fig.\ref{fig:0}c. 
The same fgure shows that on the contrary atomic decay $\gamma$ is almost irrelevant (gray dotted and dashed lines) as long as the population in $P_{\ket{\Phi_0}}$ remains small. Actually $e-g$ processes are even less relevant since these states are practically not populated and neglecting it would not make any appreciable difference. 

In the same figure we show that population transfer may also occur due to a stray corotant coupling of the cavity to the $u\!-\!g$ transition, 
$H_{ug}= \lambda^\prime \big[\ketbra{u}{g} \,a^\dagger + \ketbra{g}{u}\, a \big]$, while $\lambda=0$. As shown in Ref.~\cite{ka:19-falci-scirep-usc} in the presence of the stray coupling detecting two photons is not anymore a smoking-gun for USC in available AA-based quantum device, spoiling the experimental significance of all the proposals based on transitions in the $\Lambda$ scheme~\cite{ka:13-stassisavasta-prl-USCSEP,ka:14-huanglaw-pra-uscraman,ka:17-falci-fortphys-fqmt,ka:17-distefanosavastanori-njp-stimemission}. On the other hand  Fig.\ref{fig:0}c shows that if stray processes could be somehow suppressed there is room for unambiguous detection of USC-ground state virtual photons also for relatively large $\kappa/\omega_c \gtrsim 10^{-3}$.   

\section{The Vee configuration}

To circumvent the problem of stray couplings an alternative control scheme operating in the Vee configuration has been introduced in Ref.~\cite{ka:19-falci-scirep-usc}. This work shows that remarkably stray couplings in this case play no role. Moreover control fields may be coupled more easily to the relevant transitions allowing faster operations. On the other hand noise acts in a different way thus also in this case the regime of partial coherence has to be investigated in detail. In the Vee scheme the lower doublet of the artificial atom is coupled to the mode, while the ancillary level has higher energy. The Hamiltonian is the same Eq.(\ref{eq:H-rabi-model}) but with $-\epsilon^\prime \gg \epsilon > 0$. Population transfer occurs via the intermediate levels $\ket{\Phi_{1 \pm}}$ (the first excited doublet of the Rabi model). 
The open quantum system dynamics for Vee-STIRAP, is studied by the Lindblad Equation Eq.(\ref{eq:lindblad-lambda}), but with the control Hamiltonian $H_{\rm C}^{\Lambda}(t) \to H_{\rm C}^{V}(t) = W(t) (|u\rangle\langle e| + |e\rangle\langle u|)$ where $W(t)$ is a two-tone signal  with $\omega_p \approx \varepsilon_u-E_{1 \pm}$ and $\omega_s \approx \omega_p +2 \omega_c$. Population transfer occurs only if $\braket{\Phi_{1 \pm}}{H_c|2 u}$ is large enough, being nonzero only if the intermediate states contain virtual photons. The minimal atomic dissipator is now 
${\cal L}_{\rm atom}[\rho] = 
\frac{\gamma}{2} (2 \sigma_{\rm eu} \rho \sigma_{\rm eu}^\dagger - \sigma_{\rm uu} \rho - \rho \sigma_{\rm uu})
+ \frac{\gamma}{2} (2 \sigma_{\rm ge} \rho \sigma_{\rm ge}^\dagger - \sigma_{\rm ee} \rho - \rho \sigma_{\rm ee})
$ 
where 
$\sigma_{\rm eu}=| e \rangle \langle u| $, $\sigma_{\rm uu}=| u \rangle \langle u|$ 
and  $\sigma_{\rm ge}=| g \rangle \langle e| $, $\sigma_{\rm ee}=| e \rangle \langle e|$. 

\begin{figure}[t!]\centering
\resizebox{0.8\columnwidth}{!}{\includegraphics{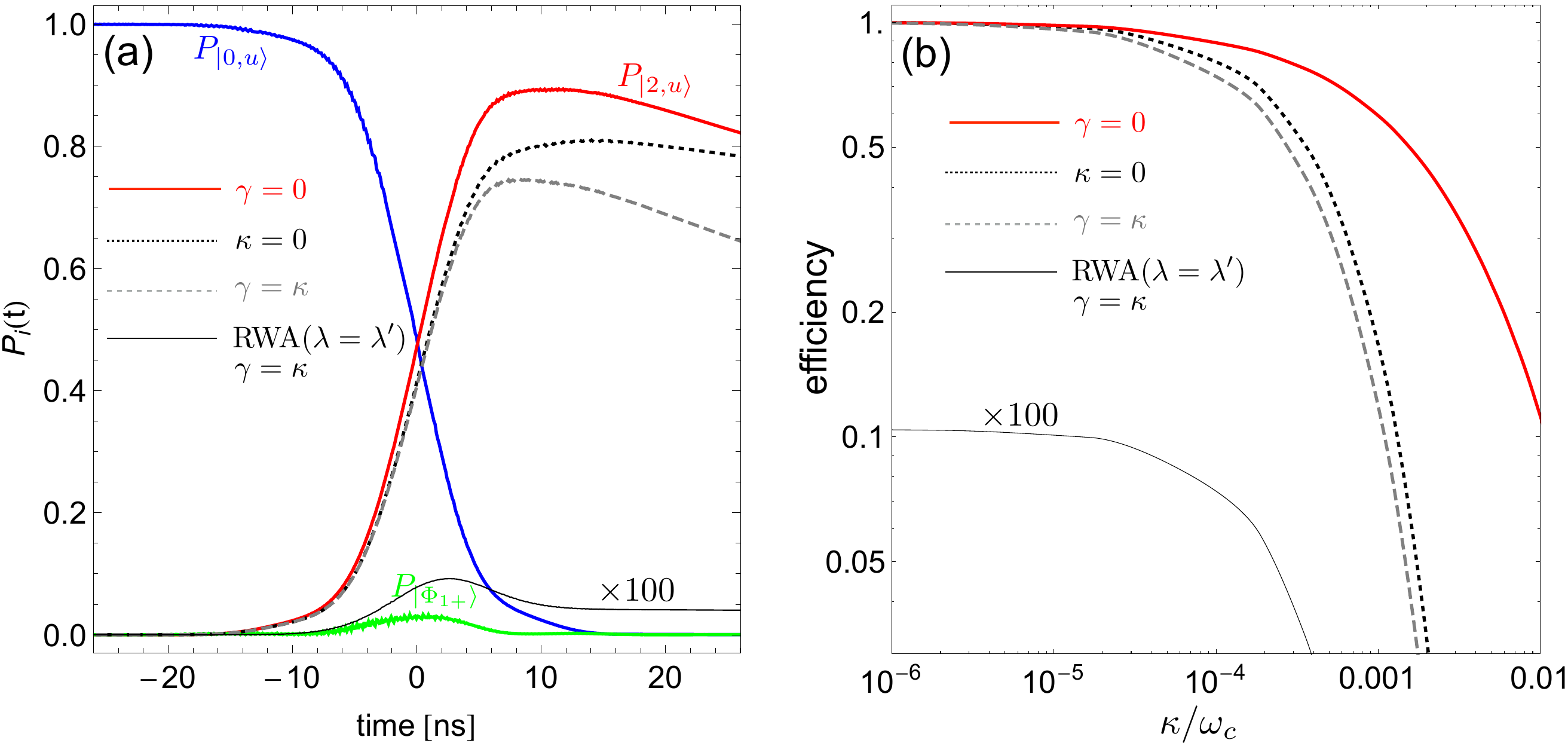}}
\caption{(a): population histories for Vee STIRAP calculated for $\lambda/\omega_{c} = 0.5$, ${\cal W}_{s}/\omega_{c} = 0.1$, ${\cal W}_{p}/{\cal W}_{s} = 0.4078$, $\kappa/\omega_{c} = 10^{-4}$, $\gamma=0$, $\epsilon^{\prime} = 1.5\epsilon$, $\epsilon = \omega_{c} = 6 \, \text{GHz}$, $T = 13.0 \, \text{ns}$ and $\tau = 0.6 T$. The $|P_{2,u}\rangle_{t}$ is calculated for $\kappa = 0$ and $\gamma/\omega_{c} = 10^{-4}$ (black dotted), $\gamma = \kappa =  10^{-4} \omega_{c}$ (gray-dashed), showing that the impact of the two main decay channels is comparable. 
The effect of the stray coupling $\lambda^\prime = \lambda =0.5 \, \omega_{c}$, but with no counterrotating term is also shown for 
$\gamma=\kappa=10^{-4}\omega_{c}$ with $\lambda = \lambda^{\prime} =0.5 \, \omega_{c}$ (black solid thin). (b): efficiency of the protocol as function of $\kappa$ with the same parameters of (a).}
\label{fig:3}    
\end{figure}

The results in Fig.~\ref{fig:3} show that in this case the atomic transition rate $\gamma$ has in general an impact even larger than the cavity losses $\kappa$. This is evident both from the population histories, Fig.~\ref{fig:3}a, and the maximal efficiency
Fig.~\ref{fig:3}b (dashed and dotted lines vs the red line).  
On the other hand the trade-off of this drawback with the possibility of faster processes (notice the time scale in Fig.~\ref{fig:3}a) 
is favorable, showing that Vee STIRAP is also robust against decoherence. Indeed in Fig.~\ref{fig:3}b, where the efficiency of the protocol is reported at the maximum value obtained during the evolution, displays figures of $\sim 50 \%$ up to a decay rate $\kappa/\omega_{c} \gtrsim 2 \times 10^{-3}$. Also in this case  decay of level $e \to g$ is negligible since $\ket{e}$ does not get populated during the dynamics owing to the high efficiency of Vee-STIRAP. It is finally shown that the effect of stray corotating couplings is irrelevant also in the presence of new channels opened by dissipation, this making  Vee STIRAP a unique tool for the dynamical detection of virtual photons in the USC regime. 

\section{Conclusions}
We studied a theoretical proposal of dynamical detection of virtual photons in the highly entangled eigenstates of an AA coupled with a quantized oscillating mode in the USC regime. Our proposal leverages on the coherent amplification of photon pair production obtained in the ideal case of a closed system by operating a control scheme similar to STIRAP. We show that detection of photon pairs is also possible with partial coherence only. In particular by driving the system in $\Lambda$ configuration the most relevant decoherence mechanism are photon losses of the cavity. The fundamental difficulty for this protocol is that in the so far available hardware the unavoidable presence of stray couplings between the cavity and unwanted transitions in the atom may yield pair production also in the absence of USC. 
Our results show that if this problem is mitigated by new design or control solutions, then decoherence does not hinder the detection of virtual photons. On the other hand the problem of stray couplings can be circumvented by resorting to STIRAP in the Vee configuration. 
We have shown that in this case the stray channels are suppressed also in the presence of decay processes. Moreover,     
while the contribution of atomic spontaneous decay has also a direct impact in lowering the efficiency, operations are faster the trade-off between the two effects being positive. This opens realistic perspectives to the detection of virtual USC photons in state of the art solid-state architectures. 

\begin{acknowledgement} 
This work was supported by the QuantERA grant SiUCs by CNR GA nr. 731473 QuantERA, and 
by University of Catania, Piano per la Ricerca 2016-18 - linea di intervento \textquote{Chance},  
Piano di Incentivi per la Ricerca di Ateneo 2020/2022, proposal Q-ICT.
\end{acknowledgement}


\begin{thebibliography}{10}

\bibitem{kb:06-harocheraimond}
S. Haroche and J.M. Raimond, 
\newblock {\em Exploring the Quantum: atoms, cavities and photons}.
\newblock (Oxford University Press, 2006)

\bibitem{kr:08-schoelkopf-nature-wiring}
R.~J. Schoelkopf and S.~M. Girvin, 
\newblock { Nature}, {\bf 451}, 664 (2008)

\bibitem{ka:04-wallraff-superqubit}
A.~Wallraff, D.~I. Schuster, A.~Blais, L.~Frunzio, R.-S. Huang, J.~Majer,
  S.~Kumar, S.M. Girvin, and R.J. Schoelkopf,
\newblock { Nature}, {\bf 421}, 162--167 (2004)

\bibitem{ka:18-distefanopaternostro-prb-measurethermo}
P.~G. Di~Stefano, J.~J. Alonso, E.~Lutz, G.~Falci, and M.~Paternostro,
\newblock Phys. Rev. B {\bf 98}, 144514 (2018).

\bibitem{ka:09-benentifalci-prl-memorych}
G. Benenti, A. D'Arrigo, and G. Falci,
\newblock { Phys. Rev. Lett.}, {\bf 103}, 020502 (2009)

\bibitem{ka:137-rabi-pr-rabimodel}
I.~I. Rabi,
\newblock { Phys. Rev.}, {\bf  51}, 652--654 (1937).

\bibitem{ka:05-ciuti-prb-intersubbandpolariton}
C. Ciuti, G. Bastard, and I. Carusotto,
\newblock { Phys. Rev. B}, {\bf 72}, 115303 (2005)

\bibitem{kr:19-forndiazsolano-rmp-usc}
P.~Forn-Diaz, L.~Lamata, E.~Rico, J.~Kono, and E.~Solano,
\newblock { Rev. Mod. Phys.} {\bf 91}, 025005 (2019)

\bibitem{kr:19-kokchumnori-natrevphys-usc}
A.F. Kockum, A.~Miranowicz, S.~De~Liberato, S.~Savasta, and F.~Nori.
\newblock { Nat. Rev. Phys.} {\bf  1}, 19--40 (2019)

\bibitem{ka:12-ridolfo-prl-photonblock}
A.~Ridolfo, M.~Leib, S.~Savasta, and M.J. Hartmann.
\newblock { Phys. Rev. Lett.} {\bf  109}, 193602 (2012)

\bibitem{ka:13-ridolfo-prl-nonclassicalradiation}
A.~Ridolfo, S.~Savasta, and M.~J. Hartmann,
\newblock { Phys. Rev. Lett.} {\bf  110}, 163601 (2013)

\bibitem{ka:12-carusotto-prl-3levusc}
I.~Carusotto, S.~De~Liberato, D.~Gerace, and C.~Ciuti.
\newblock { Phys. Rev. A} {\bf  85}, 023805 (2012)

\bibitem{ka:13-stassisavasta-prl-USCSEP}
R.~Stassi, A.~Ridolfo, O.~Di~Stefano, M.~J. Hartmann, and S.~Savasta.
\newblock { Phys. Rev. Lett.} {\bf  110}, 243601 (2013).

\bibitem{ka:14-huanglaw-pra-uscraman}
Jin-Feng Huang and C.~K. Law,
\newblock { Phys. Rev. A} {\bf 89}, 033827 (2014)

\bibitem{ka:17-falci-fortphys-fqmt}
G.~Falci, P.G. Di~Stefano, A.~Ridolfo, A.~D'Arrigo, G.S. Paraoanu, and E.~Paladino,
\newblock { Fort. Phys.} {\bf 65}, 1600077 (2017)

\bibitem{ka:17-distefanosavastanori-njp-stimemission}
O.~Di~Stefano, R.~Stassi, L.~Garziano, A.~F. Kockum, S.~Savasta, and F.~Nori,
\newblock { New J. Phys.} {\bf  19}, 053010 (2017)

\bibitem{ka:19-falci-scirep-usc}
G.~Falci, A.~Ridolfo, P.G. Di~Stefano, and E.~Paladino, 
\newblock { Sci. Rep.} {\bf  9}, 9249 (2019)

\bibitem{19-deliberato-pra-quantumvacuum}
S. De~Liberato.
\newblock { Phys. Rev. A} {\bf  100},  031801 (2019)

\bibitem{kr:17-vitanovbergmann-rmp}
N.~V. Vitanov, A.~A. Rangelov, B.~W. Shore, and K. Bergmann.
\newblock Stimulated Raman adiabatic passage in physics, chemistry, and beyond.
\newblock { Rev. Mod. Phys.} {\bf  89}, 015006 (2017)

\bibitem{ka:09-siebrafalci-prb}
J. Siewert, T. Brandes, and G.~Falci,
\newblock { Phys. Rev. B} {\bf  79}, 024504 (2009)

\bibitem{ka:15-distefano-prb-cstirap}
P.~G. Di~Stefano, E.~Paladino, A.~D'Arrigo, and G.~Falci,
\newblock { Phys. Rev. B} {\bf  91}, 224506 (2015)

\bibitem{ka:16-kumarparaoanu-natcomm-stirap}
K.S. Kumar, A.~Veps\"al\"ainen, S.~Danilin, and G.S. Paraoanu.
5\newblock Stimulated raman adiabatic passage in a three-level superconducting  circuit.
\newblock { Nat. Comm.} {\bf  7}, 10628 (2016)

\bibitem{ka:16-xuhanzhao-natcomm-ladderstirap}
H.K. Xu, W.~Y. C.~Song, G.~M. Liu, F.~F. Xue, H.~Su, Ye~Deng, D.~N. Tian,
  Siyuan Zheng, Y.~P. Han, H.~Zhong, Yu-xi~Liu Wang, and S.~P. Zhao.
\newblock { Nature Communications} {\bf  7}, 11018 (2016)

\bibitem{ka:16-vepsalainen-photonics-squtrit}
A. Veps\"al\"ainen, S. Danilin, E. Paladino, G. Falci, and
  G.~S. Paraoanu, 
\newblock { Photonics} {\bf  3}, 62 (2016)

\bibitem{kr:14-paladino-rmp}
E.~Paladino, Y.M. Galperin, G.~Falci, and B.L. Altshuler.
\newblock { Rev. Mod. Phys.} {\bf  86}, 361--418 (2014)

\bibitem{ka:19-ridolfofalci-epj-usc}
A.~Ridolfo, G.~Falci, F.M.D. Pellegrino, and E.~Paladino.
\newblock Photon pair production by stirap in ultrastrongly coupled
  matter-radiation systems.
\newblock { Eur. Phys. J. Special Topics} {\bf  227}, 2183?2188 (2019)

\bibitem{ka:16-distefano-pra-twoplusone}
P.~G. Di~Stefano, E.~Paladino, T.~J. Pope, and G.~Falci, 
\newblock { Phys. Rev. A} {\bf  93}, 051801 (2016)

\bibitem{ka:13-falci-prb-stirapcpb}
G.~Falci, A.~La~Cognata, M.~Berritta, A.~D'Arrigo, E.~Paladino, and
  B.~Spagnolo, 
\newblock { Phys. Rev. B} {\bf  87}, 214515-1,214515-13 (2013)

\bibitem{ka:03-paladino-advssp-decoherence}
E.~Paladino, L.~Faoro, and G.~Falci,
\newblock Decoherence due to discrete noise in josephson qubits.
\newblock In {Kramer, B}, editor, {\em Adv. in Sol. State Phys.}, volume~{\bf
  43} of {\em {ADVANCES IN SOLID STATE PHYSICS}}, pages 747--762. {Deutsch Phys
  Gesell, Arbeitskreis Festkorperphys}, 2003

\bibitem{ka:08-paladino-prb-coherimp}
E.~Paladino, M.~Sassetti, G.~Falci, and U.~Weiss,
\newblock { Phys. Rev. B} {\bf 77}, 041303(R) (2008)

\bibitem{ka:10-paladino-prb-opt2qubit}
E.~Paladino, A.~Mastellone, A.~D'Arrigo, and G.~Falci, 
\newblock { Phys. Rev. B} {\bf 81}, 052502(2010)




\end{thebibliography}
\end{document}